\documentclass[pdflatex,sn-basic]{sn-jnl}

\usepackage{graphicx}%
\usepackage{multirow}%
\usepackage{amsmath,amssymb,amsfonts}%
\usepackage{amsthm}%
\usepackage{mathrsfs}%
\usepackage[title]{appendix}%
\usepackage{xcolor}%
\usepackage{textcomp}%
\usepackage{manyfoot}%
\usepackage{booktabs}%
\usepackage{algorithm}%
\usepackage{algorithmicx}%
\usepackage{algpseudocode}%
\usepackage{listings}%

\begin{document}

\title[Article Title]{Outstanding questions and future research of magnetic reconnection}

\author[1*,2]{R. Nakamura}
\author[3]{J. L. Burch}
\author[4]{J. Birn}
\author[5]{L.-J. Chen}
\author[6]{D. B. Graham}
\author[7]{F. Guo}
\author[3]{K.-J. Hwang}
\author[8]{H. Ji}
\author[6]{Y. Khotyaintsev}
\author[9]{Y.-H. Liu}
\author[10]{M. Oka}
\author[11]{D. Payne}
\author[12]{M. I. Sitnov}
\author[11]{M. Swisdak}
\author[1]{S. Zenitani}
\author[11]{J. F. Drake}
\author[3]{S. A. Fuselier}
\author[3]{K. J. Genestreti}
\author[5]{D. J. Gershman}
\author[13]{H. Hasegawa}
\author[14]{M. Hoshino}
\author[6]{C. Norgren}
\author[15]{M. A. Shay}
\author[5]{J. R. Shuster}
\author[16]{J. E. Stawarz}


\affil[1]{Space Research Institute, Austrian Academy of Sciences, Schmiedlstra{\ss}e 6, 8042 Graz, Austria}
\affil[2]{International Space Science Institute, Switzerland}
\affil[3]{Southwest Research Institute, San Antonio, Texas 78238, USA}
\affil[4]{Center for Space Plasma Physics, Space Science Institute, USA}
\affil[5]{Goddard Space Flight Center, NASA, USA}
\affil[6]{Swedish Institute of Space Physics, Uppsala, Sweden}
\affil[7]{Los Alamos National Laboratory, Los Alamos, NM 87545, USA}
\affil[8]{Department of Astrophysical Sciences, Princeton University, Princeton, New Jersey 08544, USA}
\affil[9]{Department of Physics and Astronomy, Dartmouth College, Hanover, New Hampshire 03750, USA}
\affil[10]{Space Science Laboratory, UC Berkeley, Berkeley, California 94720, USA}
\affil[11]{University of Maryland, College Park, Maryland 20742, USA}
\affil[12]{The Johns Hopkins Applied Physics Laboratory, 11100 Johns Hopkins Road, Laurel, 20723, MD, USA}
\affil[13]{Institute of Space and Astronautical Science, JAXA, Sagamihara, Japan}
\affil[14]{Department of Earth and Planetary Science, The University of Tokyo, Tokyo, 113-0033, Japan}
\affil[15]{Department of Physics and Astronomy,
University of Delaware, Newark, DE 19716, USA}
\affil[16]{Department of Mathematics, Physics, and Electrical Engineering, Northumbria University, Newcastle upon Tyne, UK}


\abstract{This short article highlights the unsolved problems of magnetic reconnection in collisionless plasma. The advanced in-situ plasma measurements and simulations enabled scientists to gain a novel understanding of magnetic reconnection.  Still,  outstanding questions remain on the complex dynamics and structures in the diffusion region, on the cross-scale and regional couplings, on the onset of magnetic reconnection, and on the details of energetics.  Future directions of the magnetic reconnection research in terms of new observations, new simulations and interdisciplinary approaches are discussed.}


\keywords{magnetic reconnection, Magnetospheric MultiScale, diffusion region, onset, cross-scale, energetics}

\maketitle

\section{Introduction}\label{sec1}
Magnetic reconnection is a fundamental energy conversion process in plasmas. While the changes in the topology of the magnetic field take place inside a small region, regions of acceleration and heating of plasma are distributed at larger scales, driving plasma transport or leading to explosive magnetic energy release on large scales such as substorms, solar flares and gamma ray bursts. With the modern space technology the Geospace is an ideal plasma laboratory to study the ground truth of how collisionless magnetic reconnection operates in nature, since plasmas and fields in action can be directly measured at high cadence. With the advanced in-situ measurement capabilities onboard the four Magnetospheric MultiScale (MMS) spacecraft  \citep{Burch2016SSR}, studies of magnetic reconnection and relevant plasma processes have been significantly advanced by resolving the electron-scale physics.
The rich studies conducted in the new MMS era motivated us to summarize the up-to-date understanding of magnetic reconnection from new observations mainly in Geospace but also in other environments as well as from theoretical studies \citep{Burch2024SSR}. 

The studies based on in-situ observations from MMS and simulation confirmed the theoretical predictions and led to a number of new discoveries within the active reconnection region under diverse plasma conditions \citep{Genestreti2024SSR}. In particular, progress is made in observations and theories related to the reconnection rate and energy conversion process \citep{Liu2024SSR}, the kinetic behavior of both the electrons and ions in the vicinity of the diffusion region \citep{Norgren2024SSR}, which suggests complex 3D processes.    
The diverse roles of the waves and turbulence in the magnetic reconnection are also among the important discoveries from the MMS observations\citep{Graham2024SSR, Stawarz2024SSR}. Some of these features have not been predicted or not been focused in theory or numerical simulations before the MMS era.

MMS with other spacecraft and with empirical and/or theoretical modeling, it allows us to gain new insights also on the macroscale consequence, including the large-scale consequence of the solar-wind magnetospheric interaction \citep{Fuselier2024SSR}  and particle acceleration \citep{Oka2023SSR}, as well as the coupling among the magnetic reconnection related processes at different scales \citep{Hwang2023SSR}. 
All these studies took benefit from the new development of the data analysis techniques \citep{Hasegawa2024SSR} and simulation/modeling schemes \citep{Shay2024SSR}, which allow direct comparison between observed and simulated velocity distribution of particles and electromagnetic signatures. 

 Recent observations throughout the entire solar system environment \citep{Drake2024SSR, Gershman2024SSR} and advanced laboratory experiments \citep{Ji2023SSR} enabled us to study different scales of magnet reconnection in different parameter regimes and deepen our understanding of the reconnection. New kinetic and fluid simulations have also significantly contributed to understanding magnetic reconnection also for astrophysical plasmas both in the collisionless and collisional regimes \citep{Guo2024SSR}.

While significant advancement has been made with these endeavors, there are still a number of unsolved questions in magnetic reconnection both in the kinetic physics as well as macro-scale consequences at different environments, within and beyond Geospace. In this short paper, we highlight several unsolved questions of magnetic reconnection and propose future research direction in the upcoming years with MMS as well as for the next decades.

\section{Unsolved problems}\label{sec2}

\subsection{Complex dynamics and structures in the diffusion region}\label{subsec_complex}

Substantial progress has been made in understanding the relation between magnetic reconnection and kinetic plasma waves \citep[e.g.,][]{Graham2024SSR}. 
These include specification of the types and locations of the waves that can develop during reconnection and identification of particle distributions that can excite the waves. However, much less is known about the effects of these waves on plasma from observations and it is likewise difficult to determine how waves can affect reconnection. In particular, an ongoing question is whether anomalous resistivity due to wave-particle interactions contributes to magnetic reconnection, for example by modifying the reconnection electric field \citep[e.g.,][]{yoo24}.  MMS was able to directly quantify anomalous resistivity associated with reconnection by resolving the changes in electron distributions and moments associated with lower hybrid waves \citep{Graham2022NatCo}. The results showed that the contributions from anomalous resistivity were small in consistent with previous theoretical and observational studies,
although significant cross-field diffusion can develop, which broadens narrow boundary layers and facilitates electron mixing. 
Further works can be done with MMS to answer the question on the role of waves in reconnection by examining also electron interaction with the higher frequency waves. 
While the current direct investigation of the wave-particle interaction using the highest resolution electron distributions is limited up to around the lower hybrid frequency, 
the wave-particle correlator technique, which has been used to compute the energy transfer between waves and particles for the whistler waves in the magnetosheath \citep{Kitamura2022NatCo}, can be applied also for reconnection current sheet to study higher frequency wave-particle interaction.

Furthermore MMS had made discoveries that had not been predicted by theory or numerical simulations.  MMS observations have shown that the agyrotropic electron distributions found in the electron diffusion region can become unstable to large-amplitude waves \citep{Graham2024SSR} such as the upper hybrid waves and the electron Bernstein waves due to beam-plasma interactions.  
These waves provide potential sources of radio emission and can modify the electron distributions in the EDR, but the overall impact of these processes on the reconnection remains to be quantified. These observations also clearly demonstrate the presence of physical processes at scales below the electron gyroscale, i.e. down to Debye scale, inside the EDR. The proper description of the EDR physics must include therefore Debye-scale processes, which are not currently resolved in typical simulations (see Sec.~\ref{future_model}).

MMS have also shown  that some EDRs exhibit turbulent structures \citep{Khotyaintsev2020PRL}  or strong oscillations \cite{Cozzani2021PRL} in and around EDR.  The oscillations were attributed to kinking of the current sheet by an electromagnetic drift wave propagating in the out-of-plane direction, suggesting that reconnection needs to be considered in three dimensions.
Kinetic simulations have shown that EDRs can become structured and turbulent when there is scale separation between the electron Debye length and electron inertial length \citep{JaraAlmonte2014}. More generally, MMS observations have reported both turbulent and more laminar EDRs at the magnetopause and in the magnetotail \citep{Liu2024SSR, Graham2024SSR}. At present, it is not fully understood why some EDRs are laminar, while others are more turbulent and structured. This raises the important question of whether more complicated EDRs are being missed in observations. Although a large number of EDRs have been identified by MMS, their identification has generally relied on predictions from kinetic simulations of laminar reconnection. Further works 
are needed to identify more complex EDRs.  Methods such as tunable algorithms~\citep[e.g.,][]{bergstedt20} or machine-learning techniques~\citep[e.g.,][]{bergstedt24} can be applied in identifying relevant magnetic structures from observational data. Statistical studies can then be performed in addition to case studies, that is dominated in the research thus far, leading toward a more comprehensive understanding of the complex EDR dynamics.

At present guide-field reconnection is not as well understood as anti-parallel reconnection. 
In particular, in the strong guide-field case electrons tend to remain strongly magnetized in the EDR. Thus, there is a reduced role of the off-diagonal pressure terms in supporting the reconnection electric field and reduced agyrotropy, which is often used to identify EDRs.
Kinetic simulations demonstrate the formation of a narrow sublayer (of intensified current density) embedded within the broader EDR region on the electron inertial scale \citep{yhliu14a}. The off-diagonal pressure term only becomes significant within this sublayer that is on the electron gyro-scale.
Additionally, a strong guide field results in the out-of-plane field-aligned electron flow around the X line. This results in electrostatic waves and turbulence developing in the EDR. The reduced role of agyrotropy and the role of electrostatic turbulence in guide-field reconnection requires further investigation. Interestingly, the same out-of-plane electron flow from magnetic reconnection in the strong guide field limit may explain some features of electron precipitation for discrete aurora \citep{KHuang22a}.

\subsection{Cross-scale dynamics and regional coupling \label{sec_cross}}

Magnetic reconnection operates under the presence of a diffusion region with dissipative electric fields which are generated in the electron diffusion region (EDR). Electron physics prevails in the EDR, while Hall physics becomes significant in the ion diffusion region (IDR).  The influence of the magnetic reconnection further extends to the macroscopic systems, such as the magnetospheric boundaries and meso-scale plasma structures in Geospace, for which ideal magnetohydrodynamics (MHD) works well for its overall description. 
Since these discrete reconnection regions around the X-line are interconnected via the exchange and transport of particles, momentum, and energy, with the macro-scale system, reconnection intrinsically possesses a multi-scale and cross-scale nature.
In-situ observations in Geospace and state-of-the-art numerical simulations have significantly advanced our understanding of the multi-scale aspects of reconnection \citep{Hwang2023SSR} taking place throughout the Geospace as highlighted in Fig. 1. 
They also revealed new questions that could potentially change the current understanding and lead to a paradigm shift.

\begin{figure}[ht]
\includegraphics[width=15cm]{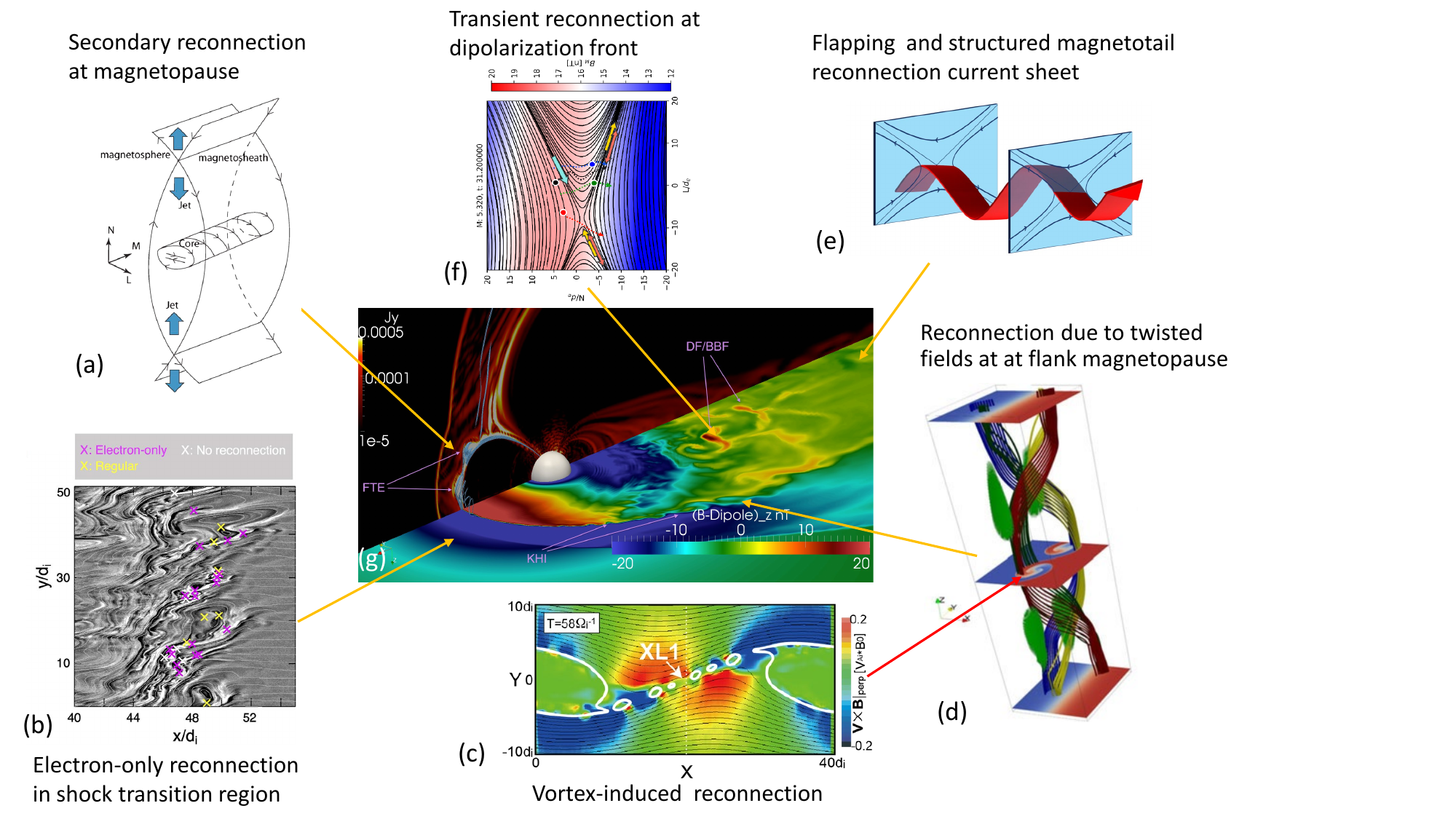} 
\caption {{\bf Reconnection in Geospace}.
Beyond the global dayside and nightside magnetic reconnection, recent in-situ measurements and simulation identified the 3D complex more localized reconnection features throughout the Geospace. (a-f) Examples of different types of reconnection that are actively studied in the MMS era. (g) the 3D view of the magnetosphere from a MHD model (Credit: V. G. Merkin) adapted from \cite{Sitnov2016EOS}, where several key mesoscale processes  KHI, BBF/DF and FTE related to localized reconnections are indicated.  The highlighed reconnection features are:   (a) secondary and/or multiple reconnection at the magnetopause (adapted from \cite{Oieroset2016GRL}, (b) turbulent reconnection in the shock transition region (adapted from \cite{Bessho2022PhP}), (c-d) 3D multi-scale KHI/KHV induced reconnection (adapted from \cite{Nakamura2011JGR} and \cite{Faganello2012EL}), and  (e) structured and disturbed EDR  in the magnetotail current sheet (adapted from \cite{Cozzani2021PRL}) and (f) transient and localized reconnection at the dipolarization front (adapted from \cite{Hosner2024JGR})}.
\label{geospace}
\end{figure}

\subsubsection{Electron-only to ion-coupled reconnection}\label{subsubsec221}

The MMS data-model analyses have provided that reconnection is ubiquitous in the shock transition region, the foreshock, and the magnetosheath downstream of both quasi-parallel and quasi-perpendicular shock (Fig.~\ref{geospace}b).  Of particular interest in this region is the newly discovered electron-only-reconnection from observations \citep{Phan2018Nat} stimulating new theoretical studies \citep{Liu2024SSR} and new investigations on interplay between turbulence and reconnection \cite{Stawarz2024SSR}.
In turbulent systems 
an electron-only reconnection is considered to occur mainly because
the scale of the turbulent fluctuations limits the maximum size of the X-line in particular along L (the main magnetic field direction in the current sheet). 
Alternatively, it has been also suggested that the electron-only reconnection might represent the early stage of regular reconnection before the X-line becomes large enough to involve ions. Such finite lifetime effects may be relevant also for magnetotail reconnection. Yet, confirming such a scenario is challenging.
It is uncertain whether the transition from electron-only reconnection to ion-coupled reconnection is regulated by the reduction of the reconnection rate,
since the rate of the electron-only reconnection was obtained to be similar to (or even higher than) the regular reconnection rate \citep{Sharma2019POP}.
Further investigation and observation are needed to gain a complete understanding of the electron-only reconnection and its role in cross-scale reconnection dynamics and possible scale-dependent energy conversion.

\subsubsection{Velocity shear driven asymmetric reconnection}\label{subsubsec223}

Asymmetry in density, velocity shear and magnetic shear are important conditions for different magnetic reconnection regimes \citep{Genestreti2024SSR}. 
Combined effects of these local asymmetries are prominent at the flank-side magnetopause, where
complex multi-scale evolution of the magnetic reconnection current sheet can take place associated with flow shear in the flank-side magnetopause developing also to turbulent layer depending on the different ambient conditions \citep{Hwang2023SSR,Stawarz2024SSR}. 
When the interplanetary field (IMF) is northward, the low-latitude magnetopause is stable to reconnection but subject to large-scale Kelvin-Helmholtz instability (KHI) driven by shear flows. Under super-Alfv\'{e}nic conditions 
the vortex flow produced by the non-linear growth of the KHI can locally compress the magnetic shear layer (current sheet), forcing the onset of
a vortex-induced reconnection (VIR) \citep{TKMNakamura17a} as shown in Fig.~\ref{geospace}c and can further develop into a complex turbulent boundary layer.
When the IMF is southward, meaning a strong magnetic shear at the magnetopause favorable for reconnection,  
the evolution of the current sheet varies depending on the initial condition (magnetic shear vs. flow shear). However, the two modes can interact with each other, leading to complex and intercorrelated dynamics. 
Understanding the interplay between reconnection and KHI (and/or Rayleigh-Taylor instability associated with density asymmetry) is important as it would control solar wind transport and energy conversion across the flankside magnetopause.  Furthermore, reconnection can also occur out of the flow shear plane due to a 3-D twist of magnetospheric and magnetosheath magnetic fields induced by Kelvin-Helmholtz vortices as shown in Fig.~\ref{geospace}d. 
which is called “mid-latitude reconnection” (MIR).
MIR occurs several Earth radii apart from the low-latitude VIR location, while being magnetically connected in 3D. Hence, the potential “communication” between the two reconnection sites can affect the solar wind transport in a complex way.
Hence the magnetic reconnection at flank-magnetopause provide an excellent laboratory for studying multi-scale (forced) 3-D reconnection.

\subsubsection{Extent and orientation of X-lines; primary and secondary X-lines}
\label{subsubsec_xlineext}

While the magnetic reconnection at the magnetopause and magnetotail are considered as the driver of the global magnetosphere circulation, observed magnetic reconnection in these large scale current sheets suggests variability in space and time and signatures of multiple reconnection \citep{Fuselier2024SSR,Hwang2023SSR}.
 One of the difficulties in interpreting in-situ reconnection events arises from the lack of information about the large-scale context of reconnection topology from observations with limited coverage. A number of unsolved questions on the temporal and spatial scales of the reconnection are therefore remained in mesoscale and large-scale context for both the magnetopause and the magnetotail.

At magnetopause the location and extent of the primary X-line is considered to be mainly determined by the global solar wind - magnetosphere interaction enabling us to predict by the maximum shear model \citep{Hasegawa2024SSR}, which is an empirical model using upstream conditions or global parameters. Yet, observations suggest transient and localized features of magnetopause reconnection, or existence and important dynamics of the multiple reconnection 
(Fig.~\ref{geospace}a). Some simulations suggest that the local physics can influence the orientation and variation of the X-line \citep[e.g.][]{Liu2018JGR}. Relationships between the primary and secondary X-lines are yet an unsolved problem. Are the secondary reconnection generated after the primary X-line  
formed by turbulence or external (e.g., magnetosheath) conditions, or a result of the departure of the X-line orientation due to local physics? 
The evolutionary path of plasmoids and flux ropes commonly generated on the dayside magnetopause via secondary/multiple X-lines are also yet to be understood. 

Although the background configuration of the magnetotail current sheet is typically 2-D and symmetric, so that the formation of a large-scale extended X-line is expected, one of the major challenges with observations is determining the extent of the reconnection region in the out-of-plane direction as reviewed in \cite{Hwang2023SSR}. 
The complex structured EDR have been identified (Fig.~\ref{geospace}e) as discussed also in Sec. \ref{subsec_complex}) suggesting finite extent of the X-line. 
While the dawn-dusk extends of bursty bulk flow (BBF)s and localized dipolarization fronts (DF) and associated localized thin current sheets suggest a finite dimension of the source, i.e. magnetic reconnection region,
these features can only be considered as indirect evidence. This is because
they could be structured also 
by the ballooning/interchange instability developed when a wider flow penetrates into the inner magnetosphere or the structured flows/DFs are created by the interchange instability itself. Furthermore, transient localized reconnection can take place also at DF (Fig.~\ref{geospace}f) so that  DF is modified as it propagates Earthward from the source region.
Yet, it is crucial to understand the extent of the reconnection region as it affects the large-scale dynamics, i.e. magnetic flux and mass transport, as well as, particle acceleration process.
The data mining tool will give us some clue on the extension of the X-line \citep{Stephens2023JGR}.  Furthermore,  the larger spacecraft separations along the MMS spacecraft orbit planned in 2024 are potentially enabling new studies of reconnection X-line in the out-of-plane direction in Earth’s magnetotail.

  \subsection{Onset of reconnection}\label{subsec21}

While the free energy of reconnection is determined by large-scale current sheet processes and its consequences affect the large-area in space, the dissipation of a tearing mode occurs at scales of the ion or electron gyroradius.  Hence the onset problem is also naturally a multi-scale problem and so far less explored area in reconnection physics. The limitation in the current observation capabilities covering all the necessary scales makes it very difficult to compare with theoretical/numerical descriptions. Here we highlight the onset problems of different types of current sheets including magnetotail, solar flare, magnetopause and other transient current sheets.

\subsubsection{Reconnection onset in Earth's magnetotail}

 For the onset problem of the near-Earth magnetotail reconnection one needs to understand both the buildup of the thin current sheet and explosive energy release. 
The observed thin current sheets are generally embedded in a thicker plasma sheet 
 with anisotropy and agytropy both in ions and electrons 
 and contain radial or azimuthal gradients 
 \citep[][and references therein]{Runov_2021JASTP}.  
 Detection of formation and evolution of thin current sheet from in-situ observation is still limited due to the sparse dataset.  The current best approach to obtain large-scale current sheet structure is data-mining method  \citep{Sitnov2019JGR}, which succeeded to predict the location of the X-line \citep{Stephens2023JGR} where the EDR/IDR were observed by MMS.  

MHD models suggested that thin current sheets are created due to deformation of the high-latitude magnetopause boundary by the reconnected and transported magnetic flux from the dayside \citep{Birn2002JGR} or due to depletion of the closed magnetic flux at the near-Earth current sheet transported toward dayside \citep{Hsieh2014JGR}. While the basic concept of the former effect obtained in the isotropic plasma description of MHD models was verified by the 2D PIC simulation \citep{Hesse2001EPS}, modeling of the onset current sheet with very small, but still finite Bz (normal component to the current sheet), where anisotropy and agytropic pressure contribution plays a role, 
is still challenging in particular to match the observations.  The mechanism leading to the onset of magnetotail reconnection with finite Bz has been extensively studied by simulations, which revealed two primary onset mechanisms  \citep[][and references therein]{Sitnov2019SSR}. The first is the electron tearing instability preceded by an external driving of the current sheet as described above to form an electron scale current sheet  \citep[e.g.][]{Hesse2001EPS,Liu2014JGR} and the second is a magnetic flux release instability in an ion-scale current sheet with a Bz hump \citep{Sitnov2010GRL}.  The latter may involve both ideal-MHD regimes 
and the ion tearing instability. 
Yet, both ion and electron tearing simulations show that the new X-lines form just $\sim$15 di ($<$2Re) from the left boundary of the simulation box, far closer to Earth than almost all observations of tail reconnection. Recently a new class of current sheets have been explored \citep{Sitnov2022JGRA} that utilize weak anisotropy to extend current sheets much further than corresponding Harris-like current sheets. The new ``overstretched ion-scale current sheets'' are agyrotropic and are supported by the off-diagonal pressure originating from Speiser ions \citep{Arnold2023GRL}. 
Yet, comprehensive stability theory for these new current sheets have yet to be developed and simulations of reconnection onset are still an active area of research.

Using in-situ observations to detect the reconnection onset is another challenge. Recent PIC simulation suggested possible observable onset features is the slightly agytropic electron distribution \citep{Spinnangr2022GRL}. But so far there is no MMS observations within less than 10 ion gyro time from onset in the vicinity of EDR exist to confirm such prediction.  Nonetheless several MMS electron observations are interpreted to be precursor of the larger scale reconnection onset based on prediction from the simulation.  These include: observation of thin electron scale current sheet with slow electron flows \citep{Wang2018GRL}; 
Divergent electron velocity flow observation without magnetic topology change \citep{Motoba2022JGR}; 
Observation of electron-scale islands in the vicinity (or as a consequence of the formation) of a major X line \citep{Genestreti2023JGR}.
Yet all these observations are snapshots of some stage of reconnection evolution predicted by some simulations. Multi-scale observations, which monitors both the ion- and electron-scale evolution of the current sheet simultaneously, are essential for confirmation of the different onset mechanism of fast reconnection in the magnetotail current sheet. 

\subsubsection{Reconnection onset in solar flares}\label{subsubsec235} 

The mechanisms of the flare onset and associated particle accelerations are also a research area with outstanding questions \citep{Drake2024SSR}.  
Similar to the magnetotail reconnection, how the magnetic energy is build up 
and how its sudden release is triggered need to be explained to understand the flare onset. The large-scale accumulation of energy preceding the reconnection onset and its transport down to kinetic length scales are important for solar flares in coronal loops, and hence it is a multi-scale problem. While the kinetic scales are inaccessible from observations, complex 3D evolution of the flare has been extensively studied based on multi-wavelength observations as well as from the in-situ measurements of the remote observation of accelerated particles. Theories for magnetic reconnection onset in the flares, such as the breakout \citep{Antiochos1999ApJ} and tether-cutting \citep{Jiang2021NatAs}, have been successful in producing the standard eruptive morphology such as a twisted CME flux rope escaping at high speed and fast reconnection in the flare current sheet below the flux rope.  Kink instability of the flux ropes in the solar corona \citep{Torok2005ApJ}, on the other hand, has been also suggested to be important for the flare onset reproducing the above eruption.  Yet, it is not established definitively from observations as well as simulations whether Alfv\'{e}nic motions cause the onset and drive reconnection or vice versa \citep{Drake2024SSR}.  Furthermore, the observed precursor local activities such as the preflare-heating and its role in the subsequent eruption are further to be understood \citep[e.g.][]{BattagliaM_2019, HudsonHS_2021}.

In contrast to the near 2D geometry magnetotail current sheet, the guide field plays a crucial role in the evolution of the reconnection current sheet in solar flare cases. 
In the presence of a strong guide field, the thermal pressure of the current sheet can play only a minor role in the force balance, since the guide field contributes to magnetic pressure at the center of the reversal and mitigates the collapse of the converging fields \citep{Leake2020ApJ,Dahlin2022ApJ}.  It is also possible for a current sheet with small finite guide field to evolve toward a "mixed" equilibrium, where the current sheet relaxation process leads to local guide field amplification \citep{Yoon2023NatCo}. The amplification of the guide field enhances the previously negligible magnetic pressure, and creates a condition where both the thermal pressure and the magnetic pressure play a significant role stabilizing the current sheet \citep{Yoon2023NatCo}.  A similar guide field amplification process has been seen in 3D MHD simulations that demonstrate a local accumulation of magnetic shear followed by outward expansion to form a thin current sheet right before the onset of a solar flare, after which the strong guide field quickly decreases by more than an order of magnitude \citep{Dahlin2022ApJ}.  Strong magnetic shear has also been associated with larger and more rapid increases in ion kinetic and thermal energy after reconnection onset in the corona, making it a potential candidate to explain the switch-on nature of solar flares \citep{Leake2020ApJ}. 
What role do other instabilities such as kink instability versus reconnection play in the flare onset is still an open question  \citep{Drake2024SSR}.

The dynamics of reconnection in the flare current sheet will span an enormous range of scales in a much complex geometry than the magnetotail. 
In 
a collisional plasma with high Lundquist numbers ($\sim 10^{14}$) such as the solar corona the  Sweet-Parker current layers are 
highly unstable to the plasmoid instability \citep{Shibata2001EP&S, Loureiro2007PhP,Bhattacharjee2009PhP} well before they can reach kinetic scales so that 
the current sheet breakups has been successfully simulated with fluid models that are supporting observations 
\citep{Daldorff2022ApJ}.   In thin current sheets layers that form between flux-ropes, on the other hand, the super-Dreicer fields induce a transition to kinetic reconnection \citep{Stanier2019PhPl}, which cannot be detected from observations.   
How do the dynamics of reconnection current layers at kinetic scales couple to energy release at the macroscale is still an open question \citep{Drake2024SSR}.

\subsubsection{Reconnection onset in different forced current sheets}

\textbf{Magnetopause reconnection:} Due the continuous solar wind driver, the reconnection onset problem at the magnetopause is less related to “when?” 
but is more about  
“where" and "what conditions”.  
Important factors are the asymmetry in the density across the current sheet and the magnetic and flow shear between the two sides of the magnetopause current sheet as reviewed in \cite{Hwang2023SSR} and \cite{Fuselier2024SSR} for Earth case and in \cite{Gershman2024SSR} for planetary magnetosphere as well as heliopause. 
The diamagnetic drift stabilization \citep{Swisdak_2010} or the shear flow-based suppression \citep{Cassak2011PhP} provide a sufficient, but not necessary condition for determining where reconnection cannot happen.
The suppression conditions has been successfully tested at Earth and planetary magnetospheres.  Yet since the Earth's magnetopause does not fulfill diamagnetic drift stabilization condition, the mechanism of determining the location of the magnetopause reconnection as well as the multiple and transient nature of the magnetopause reconnection is not fully understood (see also Sec.~\ref{sec_cross}).

\textbf{Transient forced current sheets:} 
There are a number of evidence found that local/transient thin current sheets form as a consequence of reconnection (or non-reconnection) related flows or field disturbances  \citep{Hwang2023SSR, Stawarz2024SSR} as discussed in Sec. \ref{sec_cross} and highlighted in Fig.~\ref{geospace}.  Unlike the large-scale magnetopause or magnetotail current sheets, these current sheets can be localized and/or transient and formed by dynamic processes. These include flow shear (Kelvin-Helmholtz instability) driven reconnection at the flank magnetopause \citep{TKMNakamura17a}, the shock-and turbulent driven reconnection in the magnetosheath or foreshock region
\citep{Bessho2022PhP} .  
Furthermore, the reconnection jet itself can be also a driver of the secondary reconnection 
due to colliding reconnection jet in a multi-point reconnection site \citep{Oieroset2016GRL}. In the near-Earth magnetotail transition region, 
reconnection event was found when flux rope was interacting with dipole field  \citep{Poh2019JGR}, or  at dipolarization front in the flow braking region  \citep{Marshall2020JGR, Hosner2024JGR}.  These types of reconnection are usually forced by some primary processes and the important questions are also how these primary processes create such current sheets and how these reconnection then affect the overall system. For example, important open questions for turbulence generated reconnection would be: how and how often reconnection can be generated and how such current sheet is influenced by the fluctuation characteristics, and what impact the reconnection has on the turbulence dissipation and nonlinear interactions. Exploring different regions in space with dedicated in-situ measurements may lead to further discovery of different types of thin current sheets throughout the solar system.

\subsection{Energetics, acceleration, and heating}\label{subsec2}

The energy explosively released through magnetic reconnection goes into plasma bulk flows, heating, and nonthermal particle acceleration in systems ranging from electron-scale current sheets in turbulence to the magnetospheres of accreting black holes. The nature and controlling factors of energetics in the vast array of reconnection systems are among the most compelling questions
in reconnection research. Recent development in laboratory 
\citep{Ji2023SSR},
Geospace \citep{Oka2023SSR},
solar
\citep{Drake2024SSR} and astrophysics 
\citep{Guo2024SSR} investigations present an unprecedented opportunity to establish a common framework on energetics across different systems. 
Below we list long-standing open question, and in particular, highlight how the released magnetic energy distributes between thermal and nonthermal components and between electrons and ions in the realms of magnetotail observations, solar flares, astrophysical systems, and laboratory experiments.

\subsubsection{Magnetotail observations}\label{subsubsec231}
In-situ observations in the magnetotail enable the study particle acceleration at various regions related to reconnection; e.g., diffusion region, separatrix, magnetic islands or flux ropes, outflow and dipolarization front  \citep{Oka2023SSR}. 
Distinct power law spectra for both electrons and protons are reported associated with reconnection. A puzzle is that nonthermal population is observed during quiet plasma sheet. Also, when electrons are significantly heated 
the nonthermal tail does not always become harder \citep{Oka2022PoP}. 
This is counter-intuitive because the nonthermal tail is expected to be enhanced as the temperature increases.  
 For ions, there are less studies on the energy partition between thermal and nonthermal components. A recent study suggests that ion energization is dominated by the electric field fluctuations near the ion cyclotron frequency \citep{Ergun2020ApJ}. 
 How energies are partitioned between ions and electrons is also an important unsolved problem. When ion and electron energy flux were compared in ion diffusion region of magnetotail reconnection, it was dominated by ion enthalpy, with smaller contributions from the electron enthalpy and heat flux and the ion kinetic energy flux \citep{Eastwood2013PRL}. 

One of the important factors to understand the energetics in magnetic reconnetion is the role of the turbulence in the acceleration, which was identified in the low-beta magnetotail reconnection events both for ions and electrons \citep{Ergun2020ApJ}. While the formation of the nonthermal tail distribution is generally considered based on guiding-center approximation, it remains an open question how particles interact with turbulence/waves and how they receive energization “kicks” from fluctuations which is inherently non-adiabatic interaction. It is also interesting to know how turbulence regulates the repartitioning of energy released by reconnection as a function of distance from the x-line, since energy may be transfered from the bulk outflow into the particle thermal energy or energetic particles over some distance.
Another factor affecting the energization processes in the magnetotail reconnection is the finite extent of the reconnection regions and its multiplicity as discussed in Sec.~\ref{sec_cross}. Electrons and low-energy ions, have gyroradii smaller than the typical size of the reconnection outflows and can be confined within the reconnection region. However, the heavier or energetic ions, can have the gyroradius comparable to the transverse scale of the reconnection outflow, and thus can no longer be trapped within the outflow and their acceleration may stop. For such ions to gain further increase in the energy, they need to interact with multiple reconnection events. Yet, such structures and evolution of multiple reconnection in the magnetotail is difficult to identify from the observations. A further caveat that has to be also considered in the magnetotail events is that the particle distribution observed from a spacecraft prior to an event is generally not (or not identical to) the source of the population observed afterward. For understanding the energetics of reconnection in the magnetotail, simultaneous coverage of the acceleration regions in larger context, i.e. from X-line to the outflow regions are essential. 

\subsubsection{Solar flares
}\label{subsubsec232} 
Macro-scale energy release of magnetic reconnection have been extensively observed with remote-sensing of solar fares as well as from recent in-situ measurements in the near-sun solar wind related to the interchange reconnection within the coronal holes or the reconnection in the heliospheric current sheet as reviewed by \cite{Drake2024SSR}. It is the solar flare observations that first suggested that the released magnetic energy in reconnection is partitioned into nonthermal and thermal electrons and ions. In contrast to the magnetotail reconnection, spectrum data suggest contribution of nonthermal electrons to be comparable or exceeding the thermal electrons. Significant ion energy gain are detected in the emission, although the observed emission is limited in energy range. Combining with in-situ observations of flare ejecta by Parker Solar Probe and Solar Orbiter is expected to improve our understanding of the ion energetics. 

Modeling efforts has significantly contributed to advance our understanding of the macro-scale particle acceleration mechanisms related to reconnection as summarized in \cite{Drake2024SSR}. 
Different models integrating MHD with particle descriptions have shown effectiveness in producing observed power law spectra \citep{Arnold2021PRL, Li2022ApJ}. 
These models, as they cover kinetic to large-scale MHD regimes, make it possible to compare and predict imaging spectroscopy observations of solar flares 
and the highest energy particle acceleration in astrophysical objects.  In order to make progress in understanding the energetics in the reconnection in the solar flare,  comparisons of observations with  
the predictions from these models, for instance, on the role of guide field or location of the acceleration sites, are essential. 

\subsubsection{Astrophyical systems}
In astroyphysical system, magnetic reconnection has been proposed as a mechanism to explain high-energy phenomena and radiation signatures such as pulsar wind nebulae, pulsar magnetosphere, relativistic jets, gamma-ray burst, accretion disks, and magnetars, etc \citep{Uzdensky2011SSRv,Hoshino2012,Arons2012,Guo2020}. They can take place in relativistic magnetically dominated regions in these systems. High-energy emissions are observed during reconnection as particle heating and acceleration happens, which are one of the key issues in the reconnection studies discussed in the review by \citet{Guo2024SSR}.  The relativistic reconnection events trigger acceleration in a various regime where power-law tail slope can become near unity \citep{Sironi2014,Guo2014,Werner2016,Li2023}. The direct acceleration due to reconnection electric field can also lead to power-law spectra \citep{Zenitani2001} in addition to the more common Fermi/betatron processes among the different system \citep{Guo2015,Guo2019}. Yet, the overall framework of the energy partition problem are similar to other systems and treating simultaneously the large-scale fluid behaviour and the basic particle acceleration process is a challenging problem as in other systems, considering the enormous ratio between the system size and the plasma inertial length. 
Different theories have successfully explained magnetic reconnection as a source of nonthermal particles. However, many remaining questions  (e.g., how much energy goes to thermal and nonthermal) to be understood are similar to space plasma, but in much larger spatial and temporal scales, including those observed surrounding black holes in the event horizon telescope.

\subsubsection{Laboratory reconnection energetics}\label{subsubsec233} 

With the advantage of being able to systematically quantify reconnection energetics, laboratory experiments have made substantial progress on the topic \citep{Ji2023SSR}, in coordination with numerical simulations and space observation. As the magnetic energy is converted into flows, thermal and nonthermal energization takes place at the X line, separatrices, exhausts, and far downstream. Consistent with space observation and fully kinetic simulations, the ion energy gain was found to exceed that of the electrons in laboratory reconnetion \citep{Yamada2018Nat}. Recent experiments detected directly accelerated electron by the reconnection electric field and nonthermal electrons in anti-parallel reconnection in low-beta plasmas \citep{Chien2023}, further bringing the possibility of sharing common studies with the space community. The range of system size achievable in laboratory experiments is so far within 10 ion-inertial lengths from the X line, and hence the aspects on dynamics and energy conversion at global scales are open challenges. The effects from plasma collisions need to be carefully handled for comparative studies with space plasma. Future experiments in new facilities such as FLARE \citep{Ji2022Nat} will  access both the collisional and collisionless regimes, promising fruitful comparisons with magnetic reconnection in space and astrophysical systems.


\section{Future research}\label{sec3}

Outstanding questions reviewed in the previous section motivate us to advance the current observation and computing capabilities, thinking beyond the existing framework.  Here we discuss new research aspects that can carry us farther into understanding of magnetic reconnection in nature.   

\subsection{Interdisciplinary studies}\label{subsubsec311}

The recent development of astrophysical magnetic reconnection
has strong connection with reconnection in space, solar and laboratory environments and can be extended more in future.
The development of collisionless
magnetic reconnection and kinetic simulations, starting from 1990s, laid the
solid ground for studying relativistic magnetic reconnection in astrophysics community.
It has became a common knowledge that kinetic physics supports fast magnetic
reconnection and magnetic reconnection likely leads to plasma heating
and particle acceleration \citep[][and references therein]{Birn2012SSR}.
Meanwhile, the development of relativistic magnetic reconnection %
led to new knowledge and motivations on reconnection physics and particle
acceleration mechanisms applicable to non-relativistic regime. For example, recent progress of 
theories of
reconnection rate was initiated by studies of relativistic magnetic reconnection
 \citep{Liu2017PRL}. The development of nonthermal power-law acceleration
in relativistic magnetic reconnection cleared out the doubt on whether the particle spectrum form
the formation of power-law in non-relativistic studies \citep{Guo2024SSR}. Motivated by these, particle power-law
distributions are recently achieved in non-relativistic studies  \citep{Arnold2021PRL, Li2019ApJ, Zhang2021, Zhang2024}. 
Such connection and
communication between different communities should continue and discussions should be strongly encouraged. 

Through the common framework of theory and simulations, processes occurring in solar and astrophysical systems mainly captured with large-scale remote-sensing images can be bridged to those in space and laboratory environments where plasma are ``directly" measured. 
The understanding and knowledge gained from in-situ kinetic-scale measurements in Geospace and laboratory can be applied to other planetary environments and serve as a foundation to understand larger scale systems such as solar flares and astrophysical phenomena, for example, relativistic jets in quasars. Direct comparison of the energy spectra between the solar flare and magnetotail reconnection has proven to be a successful scheme for studying particle acceleration in magnetic reconnection \citep{OkaM_2023, Drake2024SSR}. The efficiency of the reconnection in the solar wind - planetary interaction using the common frame work throughout the solar system \citep{Fuselier2024SSR,Gershman2024SSR} can serve as a reference to other stellar systems. The 3D dynamics and evolution of reconnection current sheet detected from in-situ measurements \citep{Hwang2023SSR} as well as in the controlled laboratory settings \citep{Ji2023SSR} can benefit from the knowledge of larger scale context gained from solar flare studies \citep{Drake2024SSR} and vice versa. That is, for identifying the energy conversion site and its dynamics in the solar context one can take into accout knowledge from in-situ observations by MMS \citep{Genestreti2024SSR,Liu2024SSR,Norgren2024SSR,Graham2024SSR,Stawarz2024SSR}. Communications between communities of different skill sets  are essential.

\subsection{Multi-scale observations}\label{subsubsec32}

As outlined in Section \ref{sec_cross},  cross-scale dynamics and regional coupling remains as a challenging, unsolved problem. While ion-scale and electron-scale physics have been studied by multi-spacecraft missions such as  Cluster and MMS, respectively, and THEMIS enabled also larger scale evolution, it is necessary to have a larger number of spacecraft over a wide range of scales. In this regard, various future mission concepts has been proposed such as {\it Plasma Observatory} \citep{Retino2022ExA} to cover simultaneously the ion and fluid scale at different magnetosphere boundaries, and multipoint observations with sufficient energy range to study Earth magnetotail reconnection including the larger context, such as {\it MagneToRE} \citep{Maruca2021FrASS}, {\it MagCon} \citep{Kepko2023AAS} and {\it WEDGE}  \citep{Turner2023AAS}.

It would also be interesting to study far downtail because magnetic reconnection signatures have been identified in the distant tail ($|X| \sim$ 100-200 $R_E$). While a future multi-spacecraft mission  {\it HelioSwarm} \citep{Klein2023SSR} for studying mainly solar wind turbulence also crosses downtail to $X \sim$ -60 $R_E$, it is important to push further downtail beyond this distance. 
Such an extension would allow us to study more large scale reconnection signatures, including chains of plasmoids and enable some comparison with solar flares. Note that the ion kinetic scale in the magnetotail is on the order of 100-1000 km whereas it is only 1 m in the solar corona. 

Improving solar flare observations is also crucial for facilitating interdisciplinary and comparative studies. In the next few years, {\it Solar-C} \citep{ShimizuT_2020} and {\it MUSE} \citep{CheungMCM_2022} missions will be launched. These missions will study reconnection-related phenomena by conducting spectroscopic observations in EUV wavelength with a wide and seamless temperature coverage (1-1000 eV) and with high temporal and spatial resolutions. 
However, in order to understand the energetics and fast-varying plasma processes such as shocks and reconnection, it is also important to conduct imaging-spectroscopy using X-rays \citep[e.g.][]{OkaM_2023, GlesenerL_2023}. Unlike EUV emissions that can be delayed due to ionization and recombination processes \citep[e.g.][]{ImadaS_2011_NonEq, ShenC_2013}, X-ray continua are produced via Bremsstrahlung emission without any delay. Recent advancements in the photon-counting technique and improved focusing optics are likely to cover large dynamic range at high temporal and spatial resolutions, and therefore a high-precision imaging-spectroscopy of reconnection-related phenomena is expected to be realized. The energy spectrum would be obtained seamlessly from thermal to nonthermal energy ranges, which is a crucial step toward a better comparative study between solar and space plasmas. Currently, mission concepts such as {\it PhoENiX} \citep{NarukageN_2020} and {\it FIERCE} \citep{ShihA_2023} are being developed to achieve such imaging spectroscopy using X-rays.

\subsection{Future modeling}\label{future_model}

Nowadays, modeling of magnetic reconnection largely relies on numerical simulations as presented in \cite{Shay2024SSR}. In particle-in-cell (PIC) simulations artificial parameters are often used such as the mass ratio ($m_i / m_e$) and the ratio of the plasma frequency to the electron cyclotron frequency ($\omega_{pe} / \omega_{ce}$) to reduce the computational cost. At this point, there is no consensus on how realistic these parameters should be to give physically meaningful results. Yet, these parameters need to be chosen carefully since the artificial mass ratio controls the separation between ion-scale and electron-scale physics and modifies plasma wave properties.  Interestingly, it was reported that Debye-scale turbulence alters the electron-scale dynamics \citep{JaraAlmonte2014} when a realistic frequency-ratio parameter is used. 

A major unsolved area of research is the interaction of magnetic reconnection with both mesoscale and global scale dynamics. By ``mesoscale,'' we mean length scales much larger than the ion diffusion region but still smaller than global magnetospheric scales. Examples of such multiscale interactions are the generation and dynamics of bursty bulk flows in the magnetotail as well as the reconnection and turbulence interaction in both the magnetosheath and upstream of the Earth's bow shock.  A major issue with studying these multiscale interactions is that PIC simulations are too computationally expensive to include meso and global scales. To capture the multi-scale nature of magnetic reconnection and its interaction with larger, global scale dynamics, several novel numerical schemes are being developed.
For instance, interlocking PIC and MHD models \citep{daldorff14a,toth16a} have been developed by several research groups.  In addition, a new hybrid simulation model \textit{kglocal}, that couples particle gyrokinetics within MHD simulations for particle acceleration study, was recently proposed, as detailed in \cite{Shay2024SSR}.

Several new directions are emerging, both in software and in hardware.
Due to strong requirements of the electric power, recent supercomputers have begun to use ``accelerators'' including graphic processing units (GPUs).
Since the programming model is different, 
it is often necessary to develop GPU variants of simulation codes.  
Yet, a growing number of simulation codes have been recently developed for GPUs, by overcoming the issue of yet-to-be-improved software development environment.
Another new directions are machine learning (ML) or artificial intelligence (AI) technologies \citep{Camporeale12024}. ML/AI is useful not only for post-processing the simulation data, but also for predicting solutions for our physics problems \citep{Raissi2019_PINN,Karniadakis2021_PINN}.
Furtheremore, quantum computers could be a game changer \citep{Grumbling2019_quantum}, although the timeline for creation of practical hardware for simulations is still unknown. They may allow us to calculate by far the larger number of variables than classical computers. Yet, since basic principles and logic circuits are very different, development of algorithms for simulations is required from scratch. In the next decade, when algorithms and hardware are further progressed, it is expected to become more clear whether quantum computing is promising for plasma simulations.

\section{Conclusions}\label{sec4}
The recent advancement in the in-situ plasma measurements, which enabled to study the collisionless magnetic reconnection physic including the kinetic physics, led new discoveries as well as many open questions discussed in the previous sections.  While they mainly deal with examples from Geospace, many of these open questions are applicable also to other systems including other planets, astrophysical system, and laboratory. 
Yet, in-situ measurements are limited by a specific range of plasma parameters (from location where spacecraft can fly) and a specific scales. Remote observations, on the other hand, are usually covering large-scale context of magnetic reconnection but for a limited energy range and limited resolution not covering micro-scale. Future observational capabilities tackling the multi-scale problems of magnetic reconnection are desired.  
In the MMS era the advancement of simulations also opened up a new possibility of close comparison between the observation and simulations on different scales. Applying these simulations that are "validated" by comparing with in-situ measurement, to other system in different parameter regimes using next-generation computing techniques is expected to further advance our understanding the physics of magnetic reconnection.

\backmatter

\bmhead{Acknowledgments}
The authors gratefully acknowledge all the contributions from the participants of the International Space Science Institute (ISSI) Workshop on \textit{Magnetic Reconnection: Explosive Energy Conversion in Space Plasmas}, Bern, Switzerland, June 27 - July 1, 2022. We thank the ISSI and its staff for hosting and supporting the workshop. This work is supported by the Austrian Science Fund (FWF): P32175-N27. JES is supported by the Royal Society University Research Fellowship URF/R1/201286. 

\bigskip
\begin{flushleft}%

\end{flushleft}

\bibliography{sn-bibliography}

\end{document}